\documentclass[prl,twocolumn,superscriptaddress,showpacs,preprintnumbers,nofootinbib,amsmath,amssymb]{revtex4} 

\usepackage{epsfig}
\usepackage{mathrsfs}

\begin{document}
 
\title{Bounding $W$-$W'$ mixing with spin asymmetries at RHIC}
\author{Dani\"el Boer}
\email{D.Boer@rug.nl}
\affiliation{Theory Group, KVI, University of Groningen,
Zernikelaan 25, 9747 AA Groningen, The Netherlands}
\author{ Wilco J. den Dunnen}
\email{wdunnen@few.vu.nl}
\affiliation{Department of Physics and Astronomy, 
Vrije Universiteit Amsterdam, 
De Boelelaan 1081, 1081 HV Amsterdam, The Netherlands}

\date{\today}

\begin{abstract}
The $W$ boson can obtain a small right-handed coupling to quarks and
leptons through mixing with a hypothetical $W^\prime$ boson that
appears in many extensions of the Standard Model. Measuring or even
bounding this coupling to the light quarks is very challenging. Only
one model independent bound on the absolute value of the complex
mixing parameter has been obtained to date. Here we discuss a method
sensitive to both the real and CP-violating imaginary parts of
the coupling, independent of assumptions on the new physics, and
demonstrate quantitatively the feasibility of its measurement at
RHIC.
\end{abstract}


\maketitle

As is well-known, the observed asymmetry between matter and antimatter
in the universe requires one or more new sources of CP violation,
which is one of the main reasons why physics beyond the Standard Model
(SM) is expected. One such source can arise from a heavier version of
the $W$ boson of the weak interaction, generically called $W'$ boson,
which appears in many extensions of the SM. From experimental searches
it is known that its mass would have to be larger than at least 700
GeV \cite{Abazov:2007bs,Abazov:2008vj}. Direct searches for this
hypothetical particle thus require TeV range colliders such as
Fermilab's Tevatron or CERN's Large Hadron Collider. Although the
$0.5$ TeV center of mass energy of the proton-proton collisions at
BNL's Relativistic Heavy Ion Collider (RHIC) is too small to observe
the $W^\prime$ boson directly, it could still be probed through its
mixing with the $W$ boson. 
This possibly $CP$-violating mixing 
causes a right-handed coupling of the $W$ boson to the fermions of
which the size and $CP$-violating phase are flavor dependent and 
{\it a priori} independent of the $W^\prime$ mass. 
Neither Tevatron nor LHC will be able to set competitive,
model independent bounds on this particular coupling to the 
\emph{light quarks}, which requires accurate selection of definite 
helicity states. As will be discussed, RHIC does have the
capability to measure or bound this coupling, \emph{including its
  CP-violating part}. 
The ability to control the polarization states of the colliding protons at
RHIC offers a unique advantage that compensates for the lower
energy. It allows to filter out dominant
SM contributions to become directly sensitive to new physics
\cite{Virey2,Virey3,Rykov,Kovalenko}.  
Here we will outline the relevant observables and the
possibility to measure them at RHIC specifically.
We will leave the calculational details for a
future publication, highlighting here only certain aspects and results
in order to expedite the experimental investigation. In 2009 RHIC has
had its first polarized proton collisions at 0.5 TeV, which has
already delivered the first nonzero measurement of a parity-violating
single longitudinal spin asymmetry in $W$ production \cite{Surrow:2010tn}.
The measurements discussed here require extensive running with
transversely polarized beams, like for the planned polarized 
Drell-Yan measurements \cite{Saito,Bland}.

The process under consideration is that of $W$-boson production from the
collision of two transversely polarized protons. In the SM the
coupling of the $W$ boson to the quarks is purely of $V-A$ character, i.e.\
it couples only to left-handed quarks. If the coupling is not purely 
$V-A$, for instance due to some as yet unknown physics
beyond the SM, the cross section for the collision of two protons 
polarized transversely with respect to their momenta ceases to be 
spherically symmetric around the collision axis.
\begin{figure}[!ht]
  \includegraphics[height=2.5cm]{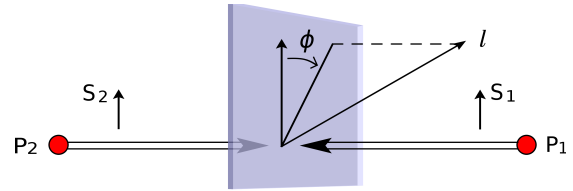}
\caption{A leptonic decay of a $W$ boson produced in a
transversely polarized proton collision. The 
transverse momentum of the outgoing lepton $l$ defines 
the azimuthal angle $\phi$ w.r.t.\ the transverse spins $S_1, S_2$ of the
colliding protons.} 
\label{Angledefplot}
\end{figure}
If the produced $W$-boson decays into an electron (or muon) and its
associated neutrino, then the electron direction can exhibit a $\cos 2\phi$
and $\sin 2\phi$ distribution w.r.t.\ the direction set by the spins,
cf.\ Fig.\ \ref{Angledefplot}. The $\sin 2\phi$ asymmetry is of particular interest
since it probes CP violation beyond the SM, as was pointed out ten years ago 
in Ref.\ \cite{Rykov}. Here we will demonstrate quantitatively the feasibility 
of measuring these asymmetric distributions at RHIC and discuss several 
essential issues, such as the required accuracy, the optimal experimental cuts,
SM background contributions, independence of assumptions on
the new physics, and the uncertainty from the transversely polarized 
quarks and antiquarks distributions. 

The reason for the asymmetries in the $\phi$ distribution is the
following. Quarks in a transversely polarized proton are also to some
extent transversely polarized, with a probability described by the
so-called transversity distribution \cite{Ralst-S-79}. A cross
section is only sensitive to transverse polarization through the
interference of left- and right-handed chirality states. Since the SM
$V-A$ coupling of the $W$ boson to the quarks only occurs for fixed
(left-handed) chirality, no sensitivity to transverse polarization
occurs in $W$-boson production \cite{Bourrely-Soffer}, except through
extremely small higher order quantum corrections. As a consequence,
double transverse spin asymmetries will be negligibly small in the SM.
Nonzero asymmetries would indicate a coupling of the $W$ boson to
right-handed quarks.  This can e.g.\ arise from the mixing with a
hypothetical $W^\prime$ boson. Such a boson arises in theories in
which a $SU(2)_\text{R}\otimes SU(2)_\text{L}$ gauge group is
spontaneously broken to $SU(2)_\text{L}$ at some scale higher than the
electroweak symmetry breaking scale. Examples are left-right symmetric
models \cite{MOHAPATRA74,SENJANOVIC75}, Little(st) Higgs models
\cite{ARKANI-HAMED02B}, SUSY $SO(10)$ \cite{BANDO94} and SUSY $E_6$
\cite{SATO96} models. Here we will consider a general model which is
not specific to any of these scenarios. It consists of a $W_\text{L}$-
and $W_\text{R}$-boson coupling to left- and right-handed particles
with strength $g_\text{L}$ and $g_\text{R}$ respectively. These states
will mix to form two mass eigenstates
\begin{equation}\label{WLWRmixing}
\begin{aligned}
W_1^{-} &= \cos{\zeta} W_\text{L}^{-} - e^{i\omega}\sin{\zeta} W_\text{R}^{-},\\
W_2^{-} &= \sin{\zeta} W_\text{L}^{-} + e^{i\omega} \cos{\zeta} W_\text{R}^{-},
\end{aligned}
\end{equation}
where $W_1$ is identified with the observed $W$ boson and $W_2$ with
the $W^\prime$ boson. Strictly speaking, new physics could lead to an
effective coupling of the SM $W$ boson to the right-handed quarks and
leptons, without the existence of a $W'$ boson. However, this scenario
is also covered by letting $M_{W_2}\to\infty$, while keeping $\zeta$
fixed. Moreover, as far as a renormalizable extension of the SM is
concerned, the $W'$ boson is the least exotic option. A nonzero
value of $\zeta$ will cause the aforementioned $\cos 2\phi$ asymmetry
to appear and if also $\omega$ is nonzero, this will reveal itself
through a $\sin 2\phi$ asymmetry.

Bounds on the mixing angle $\zeta$ are often derived by measuring the
right-handed coupling of the $W$ boson to leptons. In any process a
vanishing right-handed coupling to the leptons can result from the
right-handed neutrino, being too heavy to be produced. Therefore, it
is important to test the right-handed coupling of the $W$ boson to
leptons and quarks independently. The method discussed here measures
the right-handed coupling to quarks and is therefore independent from
the as yet unknown right-handed neutrino mass. Also, since the
coupling (including the phase) can be different for every generation
of quarks, there is no reason why that coupling in the light quark
sector should be the same as for the heavier quarks. In view of family
symmetry studies it is important to measure the couplings for all
three families separately.  Here we will focus on the light quarks,
which always suffer from additional uncertainties from nonperturbative
strong interaction effects.

The strongest bound available on $\zeta$ for quarks is, according to
the Particle Data Group \cite{PDG-GaugeAndHiggsBosons}, $\zeta <
0.003$ \cite{AQUINO91}. This bound from neutron $\beta$-decay is
obtained under a very strong assumption: manifest left-right symmetry.
This assumes Dirac-type neutrinos, an equal coupling constant for the 
left and right $SU(2)$
gauge group, equal unitary left and right CKM matrices and no complex
mixing, i.e.\ $\omega=0$. These assumptions have been questioned in
Ref.\ \cite{Langacker} and the resulting bound should not be taken at
face value. The method discussed here is independent of any of these
assumptions. The best bound available \emph{without} assumptions of
light right-handed neutrinos or manifest left-right symmetry is
$\zeta<0.04$ \cite{MISHRA92}. This has been measured in $\nu N$ deep
inelastic scattering (DIS), which is in fact the \emph{only} way in
which a model independent bound on $\zeta$ has been obtained.
Recently, there has been much discussion about the determination of
$\sin^2 \theta_W$ from $\nu N$ DIS \cite{Zeller:2001hh}, where doubts
about the employed nuclear parton densities have been raised
\cite{Eskola:2006ux,Cloet:2009qs}. Also, the strange quark can play a
significant role (cf.\ e.g.\ \cite{Ball:2009mk}), such that it
involves two generations in contrast to neutron $\beta$-decay. This
together with the fact that there is just one model independent bound
begs confirmation.

Most observables sensitive to $W$-$W'$ mixing only allow to constrain
or measure $\zeta$. The best and possibly only bound on $\omega$ can
be obtained from the bound on imaginary couplings in neutron
$\beta$-decay of Ref.\ \cite{Severijns:2006dr}. Under the assumptions
that the SM contributions do not lead to imaginary parts and that the
right-handed neutrino mass is larger than $m_n-m_p-m_e \approx 0.8$
MeV, one obtains $2\zeta\sin\omega =0.0012(19)$, which together with
the best bound on $\zeta$ translates into $\omega < 0.03$ (for $W^-$
bosons). The $\sin 2\phi$ asymmetry at RHIC will be capable of
determining or bounding the $CP$-violating phase $\omega$ for the
light quarks without these assumptions, albeit not down to such low
values. Nevertheless, it would be worthwhile to obtain an independent
bound on $\omega$, free of right-handed neutrino mass assumptions.

Now we turn to the asymmetry estimates. In lowest order in the
electroweak and strong coupling constants $\alpha$ and $\alpha_s$, the
observable under consideration becomes a product of transversely
polarized quark and antiquark distributions (denoted by $h_1^q$ and
$h_1^{\bar q}$)
convoluted with the process of quark-antiquark annihilating into a $W$
boson.  A first determination of the transversity distribution for up
and down quarks was obtained recently using semi-inclusive deep
inelastic scattering and electron-positron annihilation data
\cite{Anselmino:2007fs}. Given the considerable uncertainties in this
determination, below we will simply take $h_1^q(x)=f_1^q(x)/2$, which
is slightly above the best fit, but certainly compatible with it
within errors and is in reasonable agreement with lattice results for
the integral over the momentum fraction $x$ that requires somewhat larger $h_1$
\cite{Wakamatsu:2007nc}. Here $f_1$ denotes the unpolarized quark
distribution. From Ref.\ \cite{Martin:1997rz} it can be concluded that
for the relevant $x$-values ($x \sim 0.2$) in $W$ production at RHIC
at $0.5$ TeV, the ratio $h_1^q(x)/f_1^q(x)$ has little scale
dependence.

The absence of experimental data on the antiquark transversity
$h_1^{\bar{q}}$ prevents making absolute predictions for the
asymmetries discussed here, but for estimates we will use $h_1^{\bar
  q}(x)=f_1^{\bar q}(x)/2$, which allows for easy rescaling of the
results in the future. This choice is below its maximally allowed
value given by the Soffer bound $|h_1^{\bar q}(x)| \leq
\frac{1}{2}(f_1^{\bar q}(x)+g_1^{\bar q}(x))$, where $g_1$ denotes the helicity
distribution. The assumption $h_1^{\bar q}(x)=f_1^{\bar q}(x)/2$ may
nevertheless be an overestimate, since the scale dependence of the
ratio $h_1^{\bar q}(x)/f_1^{\bar q}(x)$ is not negligible. It decreases 
by about a factor of 2 from low energy hadronic scales to 
the relevant energy scale set by the
$W$ mass \cite{Martin:1997rz}. Fortunately, at RHIC the product
$h_1^q(x_1) h_1^{\bar q}(x_2)$ can be measured from a spin asymmetry
in the Drell-Yan process \cite{Ralst-S-79,Martin:1997rz}.
Hence the uncertainty in the asymmetry bounds below
coming from the transversity distributions can in principle be
eliminated from the analysis.

We will look at both positively and negatively charged $W$-boson
production. We restrict to their leptonic decay, which means the $W$
momentum cannot be determined.  The three independent kinematic
variables that can be measured are chosen to be the transverse
momentum of the charged lepton $l_T$, its rapidity $Y$ and the angle
$\phi$ in the plane perpendicular to the beam axis. We will not give
the full differential cross section here (cf.\ \cite{BDinprep}), but
immediately turn to the asymmetries between the processes with
parallel and antiparallel proton spins, given by the cross sections
$d\sigma^{\uparrow\uparrow}$ and $d\sigma^{\uparrow\downarrow}$,
respectively.  We define symmetric and antisymmetric cross sections as
$d\sigma \equiv
\frac{1}{2}(d\sigma^{\uparrow\uparrow}+d\sigma^{\uparrow\downarrow})$
and $\delta d\sigma \equiv
\frac{1}{2}(d\sigma^{\uparrow\uparrow}-d\sigma^{\uparrow\downarrow})$.
The latter cross section is a function of $\phi$. We define two
independent transverse spin asymmetries that select the $\cos2\phi$
and $\sin2\phi$ contributions respectively, by appropriate integration
over the azimuthal angle
\begin{equation}\label{DefAsymm}
\begin{aligned}
  A_{TT} &\equiv \frac{\left(\int_{-\pi/4}^{\pi/4} -
  \int_{\pi/4}^{3\pi/4} + \int_{3\pi/4}^{5\pi/4} -
  \int_{5\pi/4}^{7\pi/4}\right)d\phi \delta d\sigma
  }{\int_0^{2\pi}d\phi d\sigma},\\
  A_{TT}^\perp &\equiv \frac{\left(\int_{0}^{\pi/2} -
      \int_{\pi/2}^{\pi} + \int_{\pi}^{3\pi/2} -
      \int_{3\pi/2}^{2\pi}\right)d\phi \delta d\sigma }{\int_0^{2\pi}d\phi d\sigma}.
\end{aligned}
\end{equation}
The asymmetries depend on the center of mass energy and the cuts
imposed on the $Y$ and $l_T$ integrations. Transversely polarized
proton-proton collisions are only planned at RHIC, therefore, the
center of mass energy is chosen to be 0.5 TeV. The parton distribution
functions $f_1$ are taken from the CTEQ5 LO pdf set \cite{CTEQ5pdf}. 
Strange quark contributions are small and will be neglected.

The asymmetries for $W^\pm$ production (indicated by a $\pm$ 
superscript) are now given by
\begin{equation}
A_{TT}^\pm  = A^{\pm} 
\zeta_g\cos\omega, \quad \text{and,} \quad  
A_{TT}^{\perp\pm}  = B^{\pm} 
\zeta_g\sin\omega,
\end{equation}
for both beams fully transversely polarized. Here the complex phase 
$\delta_{ud}$ of the right-handed CKM-matrix element,
$V_{ud}^R=e^{i\delta_{ud}}|V_{ud}^R|$, that cannot be distinguished
from $\omega$, is absorbed into $\omega$. Also, the ratio of left and 
right coupling constants and CKM-matrix elements is conventionally 
absorbed into
$\zeta_g\equiv \zeta g_\text{R}|V_{ud}^R|/g_\text{L}|V_{ud}^L|$.
Only terms up to first
order in $\zeta_g$ are kept. 

In table \ref{tablecoefs} the values of $A^\pm$ and $B^\pm$ 
are given in leading order (LO)
approximation. The coefficient $B$ is antisymmetric in $Y$, therefore
it is calculated for \emph{half} the indicated rapidity interval. 
One can still use both forward and backward events by
taking into account this minus sign, therefore the cross section is
calculated for the \emph{full} rapidity range. The indicated range is
covered by the central detector of the STAR
experiment at RHIC. To optimize the discovery potential, i.e.\ the
ratio of the expected asymmetry to the expected statistical error, for $B$ the most central
region is excluded as it vanishes at zero rapidity. The $l_T$ range has a lower
cut off, as the asymmetry decreases at low $l_T$. The optimal values
are given in the table. At next-to-leading order the cross sections
are typically 25-40\% larger.

\begin{table}[!ht]
\centering
\begin{tabular}{l|ccc}

                        &$\quad W^+ \ $  &$\quad W^- \ $ &$\quad W^+ + W^-$\\
\hline
$A$                   	&\ -0.22	&-0.28         	&-0.23\\
$B$                   	&\ 0.16     	&-0.12         	&0.10\\
$\sigma_1[\text{pb}]\ $ &\  40        	&10           	&51\\
$\sigma_2[\text{pb}]\ $ &\  18       	&5.3       	&23\\
\end{tabular}
\caption{Coefficients and cross section at $\sqrt{s}=$0.5 TeV,
  rapidity range $|Y| \leq 1$ and transverse momentum interval $31\leq l_T \leq 45$ GeV
for $A^\pm$ and $\sigma_1$, $0.3\leq
  |Y|\leq 1$ and $35\leq l_T \leq 45$ GeV for $B^\pm$ ($Y>0$) and $\sigma_2$.}
\label{tablecoefs}
\end{table}

\begin{figure}[!ht]
  \includegraphics[width=0.3\textwidth]{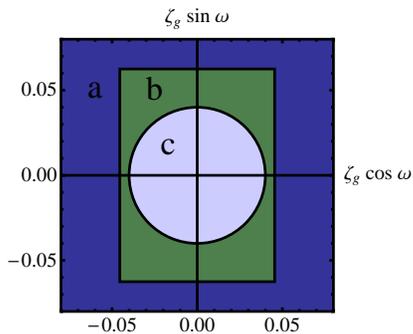}
\caption{Example exclusion plot of $\zeta_g$ and $\omega$ if the 
  asymmetries $|A_{TT}^+|$ and $|A_{TT}^{\perp+}|$ will be bounded by
  $1\%$. The region (a) would be excluded by both the best existing
  model independent bound \cite{MISHRA92} \emph{and} the new asymmetry
  measurements, region (b) would be excluded by the existing
  bound, and region (c) would be allowed by both measurements.}
\label{BoundPlot}
\end{figure}

Crucial for the possibility to measure or exclude new
physics, is the expected accuracy in the determination of the double
spin asymmetries. Translating the best model independent bound on
the right-handed coupling of the $W$ boson to the light quarks
\cite{MISHRA92}, $\zeta_g < 0.04$, into the asymmetries results in
$|A_{TT}^+|<0.9\%$ and $|A_{TT}^{\perp+}|<0.6\%$. 
If at RHIC the original design integrated luminosity of
$800\text{pb}^{-1}$ and polarizations $\mathscr{P}_1$ and
$\mathscr{P}_2$ of 70\% are achieved \cite{Saito}, 
we estimate (in agreement with \cite{NADOLSKY03}) 
the error in the spin asymmetry $\delta A_{TT} =
1/(\mathscr{P}_1\mathscr{P}_2\sqrt{\mathscr{L}\sigma})$
to be on the percent level. 
If a bound of $|A_{TT}^+|<1\%$ and $|A_{TT}^{\perp+}|<1\%$ in $W^+$
production would be obtained, 
then the bounds on the mixing become $|\zeta_g\cos\omega|< 4.5\%$ and 
$|\zeta_g\sin\omega|< 6.3\%$, showing that RHIC can deliver competitive
bounds, see Fig.\ \ref{BoundPlot}. 
Of course, the main uncertainty in these numbers comes from
the unknown magnitude of the antiquark transversity distribution,
which we emphasize can be determined simultaneously at RHIC from 
an independent asymmetry measurement. 

We end with a discussion of the expected background. Deviations from
the SM $V-A$ coupling may be generated effectively in higher orders in
$\alpha$ or $\alpha_s$, for instance by the exchange of a Higgs boson
or gluon between the annihilating $q \bar{q}$ pair. Such higher order
corrections are all suppressed by a factor of $\alpha_{(s)} m_u
m_d/M_W^2$ producing unmeasurably small asymmetries.  Higher twist QCD
corrections and partonic transverse momentum effects beyond 
collinear factorization may also generate (residual) 
double transverse spin asymmetries
within the SM \cite{Boer:2000er}, but are suppressed by at least a factor of
$M_p^2/M_W^2$ \cite{BDinprep}. 
Therefore, in the SM double transverse spin asymmetries in $W$-boson
production are at most of the $10^{-4}$ level.

We expect the largest experimental background to come from
misidentified events. This can be caused by missing a lepton from a
neutral current event interpreted as a neutrino from a charged current
event. The cross section for such a missing lepton with $|Y|>1$ is in
the order of a picobarn, leading to false $A_{TT}$ asymmetries smaller
than $10^{-3}$. For $A_{TT}^\perp$ the only neutral current
contribution comes from the interference of photon and $Z$-boson
contributions. It is proportional to the $Z$-boson width. Again
this contribution can be safely ignored.  Another type of
misidentified event can come from heavy quark decays, but this
background is largely removed together with the cuts that remove
dijet events \cite{Surrow:2010tn}.

In conclusion, without background to worry about, the double
transverse spin asymmetry in leptonic decays from $W$ bosons produced 
in polarized proton--proton collisions is a very
clean and promising way to study {\it separately} 
the mixing angle and CP-violating
phase arising from a hypothetical $W'$ boson. 
We have estimated the size of the asymmetries, without any model
dependent assumptions regarding the right-handed sector.
These estimates do depend on an assumption about the unknown distribution of 
transversely polarized antiquarks, but this can be determinded simultaneously 
through a measurement of the polarized Drell-Yan process.
We find that at RHIC, which is the only high energy polarized proton
collider, competitive bounds may be set if design goals will be
reached at 0.5 TeV. Since there is only one model independent bound on
the $W$-$W'$ mixing angle, this is a highly desirable measurement.

We thank Max Baak, Les Bland, Lex Dieperink, Vladimir Rykov, Naohito
Saito, Ernst Sichtermann, Bernd Surrow, Marco Stratmann, and Rob
Timmermans for fruitful discussions.  This work is part of the
research program of the ``Stichting voor Fundamenteel Onderzoek der
Materie (FOM)'', which is financially supported by the ``Nederlandse
Organisatie voor Wetenschappelijk Onderzoek (NWO)''.

\end{document}